\documentclass[sigconf]{acmart}

\AtBeginDocument{%
  \providecommand\BibTeX{{%
    \normalfont B\kern-0.5em{\scshape i\kern-0.25em b}\kern-0.8em\TeX}}}

\usepackage{dsfont}
\usepackage[switch]{lineno} 
\usepackage{enumitem}
\usepackage{mathtools}
\usepackage{amsmath}
\usepackage{bbm} % math font
\usepackage{amsfonts}
\usepackage{multirow}
\usepackage[ruled,vlined,linesnumbered]{algorithm2e}
\usepackage{algorithmic}
\usepackage{tabularx}
\usepackage{booktabs}
\usepackage{hyperref}
\usepackage{url}
\usepackage{fancyhdr}
\usepackage{caption}
\usepackage{subcaption} % Multiple figures

\newtheorem{definition}{Definition}[section]

\usepackage{color}
\usepackage[colorinlistoftodos]{todonotes}

\newcommand{\graph}{trajectory flow map }
\newcommand{\Graph}{Trajectory Flow Map }
\newcommand{\GETNext}{GETNext}
\newcommand{\attnmap}{transition attention map }

\copyrightyear{2022} 
\acmYear{2022} 
\setcopyright{acmcopyright}\acmConference[SIGIR '22]{Proceedings of the 45th International ACM SIGIR Conference on Research and Development in Information Retrieval}{July 11--15, 2022}{Madrid, Spain}
\acmBooktitle{Proceedings of the 45th International ACM SIGIR Conference on Research and Development in Information Retrieval (SIGIR '22), July 11--15, 2022, Madrid, Spain}
\acmPrice{15.00}
\acmDOI{10.1145/3477495.3531983}
\acmISBN{978-1-4503-8732-3/22/07}

\acmSubmissionID{692}

\begin{document}
\fancyhead{}

%% The "title" command has an optional parameter, allowing the author to define a "short title" to be used in page headers.
% enhanced/informed/aligned
% \title{GTRec: Transition Graph enhanced Transformer for Next POI Recommendation}
%\title{GTRec: Trajectory Flow enhanced Transformer for Next POI Recommendation}
% \title{GETNext: Trajectory Graph Enhanced Transformer for Next POI Recommendation}
\title{GETNext: Trajectory Flow Map Enhanced Transformer for Next POI Recommendation}
%visit graph/sequence graph/transition graph/Trajectory Graph/Trajectory Flow/Trajectory Flow Map

%% The "author" command and its associated commands are used to define the authors and their affiliations.
%% Of note is the shared affiliation of the first two authors, and the "authornote" and "authornotemark" commands used to denote shared contribution to the research.

\author{Song Yang}
\affiliation{%
  \institution{The University of Auckland}
  \city{Auckland}
  \country{New Zealand}}
\email{syan382@aucklanduni.ac.nz}

\author{Jiamou Liu}
\affiliation{%
  \institution{The University of Auckland}
  \city{Auckland}
  \country{New Zealand}}
\email{jiamou.liu@auckland.ac.nz}

\author{Kaiqi Zhao}
\affiliation{%
  \institution{The University of Auckland}
  \city{Auckland}
  \country{New Zealand}}
\email{kaiqi.zhao@auckland.ac.nz}

%% By default, the full list of authors will be used in the page headers. Often, this list is too long, and will overlap other information printed in the page headers. This command allows the author to define a more concise list of authors' names for this purpose.

\begin{abstract}
% Next POI recommendation an emerging field of location-based services, 
Next POI recommendation intends to forecast users' immediate future movements given their current status and historical information, yielding great values for both users and service providers. However, this problem is perceptibly complex because various data trends  need to be considered together. This includes the spatial locations, temporal contexts, user's preferences, etc. Most existing studies view the next POI recommendation as a sequence prediction problem while omitting the collaborative signals from other users.  Instead, we propose a user-agnostic global trajectory flow map and a novel Graph Enhanced Transformer model (GETNext) to better exploit the extensive collaborative signals for a more accurate next POI prediction, and alleviate the cold start problem in the meantime. GETNext incorporates the global transition patterns, user's general preference, spatio-temporal context, and time-aware category embeddings together into a transformer model to make the prediction of user's future moves. With this design, our model outperforms the state-of-the-art methods with a large margin and also sheds light on the cold start challenges within the spatio-temporal involved recommendation problems.
\end{abstract}




\maketitle

%\linenumbers
\section{Introduction}
%\head{Background} 
Location-Based Services (LBS) is gaining significant advancements %a revolution 
in recent years owing to the prevalence of GPS-enabled mobile devices. Service providers such as Foursquare, Gowalla, or Yelp enable users to share their experiences, tips, and moments on points of interest (POIs). Large volumes of spatio-temporal data are being accumulated, as a result, giving rise to increasingly powerful {\em next POI recommendation} systems \cite{feng2015personalized}.
%\head{Next POI rec} 
Such systems aim to predict the next POI that users will visit given their current and historical footprints (i.e., check-ins). They not only help users to better explore their surroundings but also facilitate businesses to improve their advertising strategies \cite{jiang2015author}. % to attract target customers \cite{jiang2015author}.

\begin{figure}
	\centering
	\includegraphics[width=0.9\linewidth]{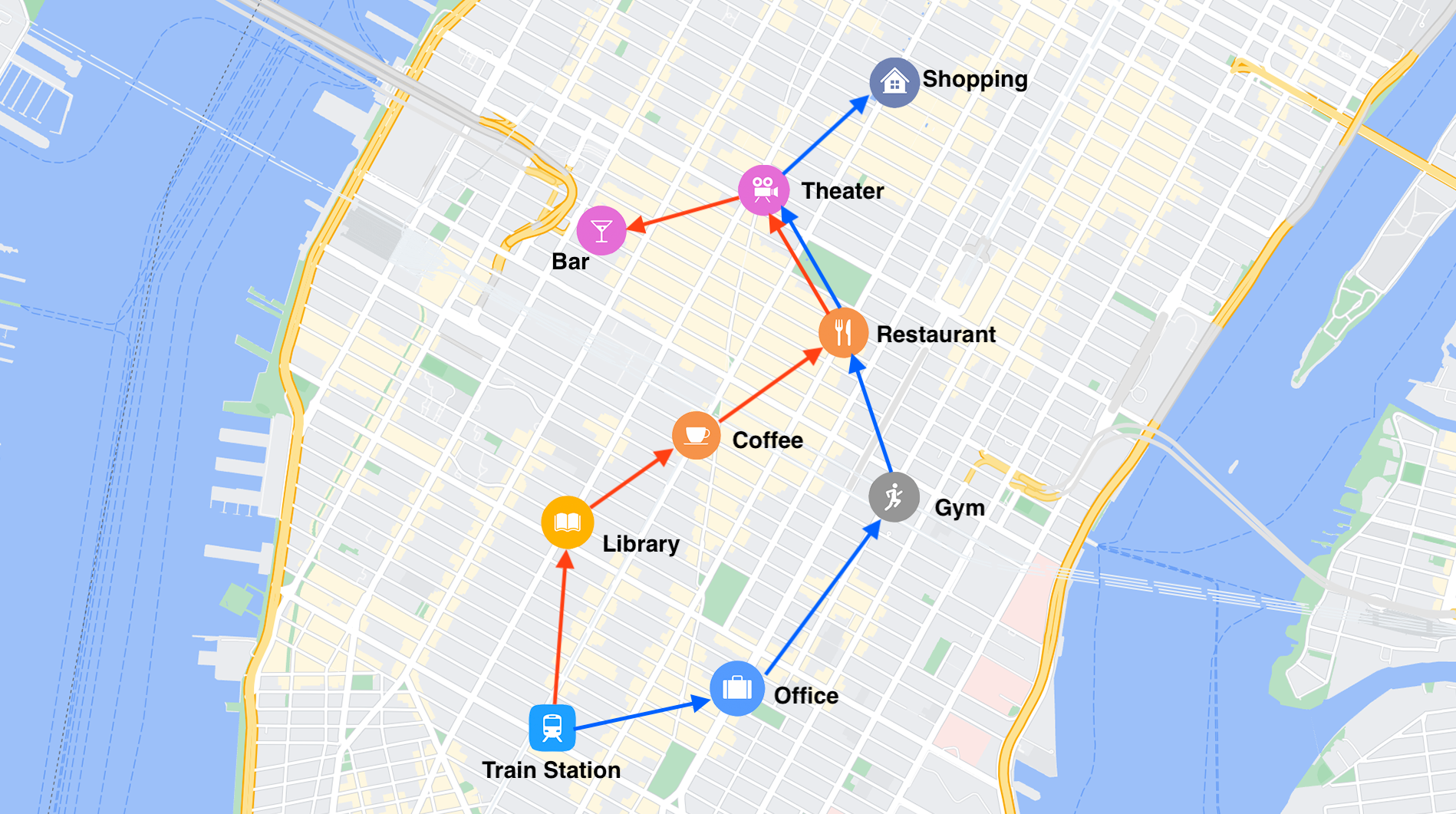}
	\caption{Two trajectories (red and blue) of two users in different days share the same fragment (restaurant to theater) in Manhattan, NYC}
	\label{fig:transit_graph}
	%\vspace{-0.3cm}
\end{figure}

As (short-term) temporal patterns of a user's trajectory provide critical insights in predicting the user's future movements, the predominant approach treats next POI recommendation as a sequence prediction task. 
As a result, state-of-the-art models for this task employ various forms of recurrent neural networks (RNN) to encode spatio-temporal information~\cite{liu2016predicting,wu2019long,zhao2020go,sun2020go,zhao2020discovering,wu2020personalized}. 
%
%However, in general POI recommendation task, such sequential information is not explicated defined. Thus, a decent number of studies construct various graphs such as user-POI heterogeneous graph, POI-region bipartite graph, user-user social network, etc and adopted sophisticated graph learning methods for POI recommendation \cite{yuan2014graph, xie2016learning,yang2019revisiting}. This huge difference makes us wonder if the graph structure data is underutilized in the next POI recommendation problem since it share the similar data with the general POI recommendation.
%
There are, however, three limitations to this approach. First, the model performance drops significantly on short trajectories when compared with long trajectories. This is due to the limited spatio-temporal contexts provided by the short trajectories. Second, the recommendation accuracy for inactive users, who have only checked in a small number of POIs, is often low. It is a commonplace that cold-start problem exists in real-life recommender systems. Third, the POI category usually presents strong temporal correlations. For example, the check-in frequency of the train station is significantly higher on rush hours compared with midnight as shown in Figure \ref{fig:time-dst}. However, existing models fail to bridge the time and POI category. 
While no silver-bullet exists in tackling the first two problems due to inherent data scarcity issues, we argue that these limitations can be mitigated to some degree by harnessing collective information from other users. In particular, individuals tend to share certain fragments of check-in sequences, forming collective {\em trajectory flows}. For instance, two users went to the same restaurant and the same theater in the same order but on different days, as shown in Figure~\ref{fig:transit_graph}. These trajectory flows may provide crucial hints on the generic movement patterns of users, helping to resolve issues with short trajectories and inactive users. Nevertheless, utilizing these trajectory flows in next POI recommendation is not straightforward, as we need to address the following questions: 
% 1): graph+GCN, 2): spatio-temporal context + time-cat, 3): global+local (Transformer and overall framework)
{\bf (1)} how to aggregate information from check-in sequences to form a unified representation of global trajectory flow patterns?
{\bf (2)} how to reserve the important spatio-temporal contextual information such as category information and user preference besides these trajectory flows? 
and {\bf (3)} how to leverage all information above in next POI recommendation, to strike a balance between generic movement patterns and personalized demands? %individual temporal features?

%这3点和后面的3段一一对应，需要在接下来的3段里呼应这三点
% These information are seldomly discussed and exploited. A possible explanation is these sharing information in next POI recommendation are not easy to utilized since the intrinsic sequential characterization. Thus, the problem is how to represent the data to exploit these similarities in trajectory. 

For {\bf Question (1)}, we construct a novel \textit{(user-agnostic) \graph} that summarizes trajectories between POIs as well as features of each POI in a single graph structure. More precisely, nodes in this graph are POIs with attributes including geographical location, category, and check-in counts. A directed edge connects a POI to another if they are successively visited in the same check-in trajectory, and edge weight represents their 
co-visit frequency. 
Different from the graphs designed for modeling the correlations between users and items in existing POI and product recommendation models, this \graph captures the transitional influence between POIs. Then
%
%
% Inspired by lift style similarity (Life style: allocate time among different types of activities) \todo{literature, background}. How to represent the data to utilize these similarities. We propose the ... graph. Future: User category (sequential pattern) -> build graph per category
%
%\head{Introduce \graph} 
%Most works in general POI or online shopping recommendation employ all sorts of graphs. However, those graphs are not suitable to be used in the next POI recommendation directly. The main reason is they usually lack of the sequential information that characterizes the next POI recommendation. Therefore, we construct a user-agnostic global \graph which preserves the global check-in transition patterns, POI background information, and visit order frequency. 
%
to exploit the collective information in the trajectory flow map, we employ a graph convolutional network (GCN) to embed POIs into a latent space that preserves the global transitions among POIs.
%and also adjust the final output directly by a learned \attnmap.
More precisely, the GCN updates the embedding of each node by aggregating from its in-neighbors' embeddings. In this way, the embedding of each POI in the \graph will be influenced by its precedents, and thus the information of global transitions are well preserved. Consequently, without knowing the historical check-ins of a user, one can still recommend the most probable out-neighbor of the current POI. An underlying assumption behind the global \graph is that recommending the popular next POI may not be the best choice but is certainly improved from a random guess. In particular, using global \graph should improve top-$k$ accuracy for relatively large $k$. See Sec.~\ref{sec:graph}.

For {\bf Question (2)}, the \graph mainly captures the user-agnostic POI transition patterns. Meanwhile, users' preferences and spatio-temporal context are essential for personalized recommendations. The long-term preferences measure a user's general tastes, such as the tendency to a favourite restaurant or a movie theater. The short-term preferences are reflected by the recent check-ins of the user, which provide a more concrete spatial-temporal context. 
%For example, if it is lunch time and current POI is office, a nearby restaurant could be the next POI recommended to the user. 
Thus, we utilize the embedding layers to capture users' general preference, POI category embeddings, and a time2vec model to depict time embeddings. Moreover, the POI categories usually show strong correlations with time. To bridging them together, 
%we firstly learn the time embedding and category embedding separately. Then, 
category and time embeddings are merged and feed into a fusion module to produce the time-aware category context embedding. See Sec.~\ref{sec:context}.

% \head{Our method} 
Summarising the above, we propose a Graph Enhanced Transformer framework GETNext \footnote{The code is available at \url{https://github.com/songyangco/GETNext}} that unifies the generic movement patterns, user's general preferences, short-term spatio-temporal contexts, and time-aware category embeddings to predict user's next POIs. Compared with RNN and LSTM, transformer can learn the contribution of each check-in directly from the input trajectory to the final recommendation using the self-attention mechanism. In other words, the model may differentiate the informativeness of different check-ins and aggregate all check-ins in the trajectory simultaneously for prediction, yielding superior performance.
%The Transformer is a purely attention based model where data can be processed in parallel and the contribution of each time step to the final prediction is learned dynamically. However, the input sequence data of RNN or LSTM needed to be processed in order since each cell depends on the output state of previous cell. 
In \GETNext\;we adopt the transformer encoder with several multi-layer perceptron (MLP) decoders to integrate the implicit global flow patterns that encoded in POI embeddings and other personalized embeddings. Moreover, the global flow patterns are also explicitly injected to the final prediction by a learned transition attention map. 
This novel architecture answers {\bf Question (3)} above. See Sec.~\ref{sec:transformer}.

To validate our proposed model, we conduct a series of experiments on well-known real-world datasets. Our experiment results show that \GETNext\;are able to  significantly outperform existing state-of-the-art methods. E.g., up to 11\% in terms of top-5 accuracy on NYC dataset. See Section~\ref{sec:experiment}. We list main contributions of our work below: 
\begin{enumerate}[leftmargin=*]
    \item We propose a global \graph to model the common visiting-order transition information, and utilize a graph convolutional network to encode them into POI embedding. 
    %encode the transition patterns as well as the geographical, meta information into POI embeddings.
    \item We develop a novel time-aware category context embedding to capture the diverse temporal patterns of POI categories. %Other papers use time directly which may corrupted by noise data for one POI. 
    \item We propose a transformer-based framework to integrate global transition patterns, user general tastes, user short term trajectory, and spatio-temporal context together for next POI recommendation.
\end{enumerate}

\section{Related Work}
% 写出每类方法的问题，突出我们的优势。research gap （呼应introduction里的challenges）

\subsection{Next POI Recommendation}
% Pioneer work \cite{cheng2013you}
Compared with conventional POI recommendation, next POI recommendation \cite{cheng2013you} focuses more on the temporal influence of the recent trajectory to predict a user's next moves.
% \paragraph{Markov Chain} 
Early studies adopt methods that have been widely used in other sequential recommendation tasks such as Markov chains \cite{cheng2013you,zhang2014lore,ye2013s}. For instance, the pioneering work by Cheng et al. \cite{cheng2013you} recommended next POIs using a matrix factorization method which embeds personalized Markov chains (FPMC) \cite{rendle2010factorizing}. Likewise, 
%Ye et al. \cite{ye2013s} developed a mixed hidden Markov model to predict the category as well as the POI, and 
Zhang et al. \cite{zhang2014lore} proposed an additive Markov chain to model the sequential transitive influence. Meanwhile, other studies explored the possibility of tailoring the commonly-used matrix factorization or metric embedding technique into the next POI recommendation \cite{feng2015personalized,zhao2016stellar,liu2016unified}.
In general, %the representation power of 
these early methods are rather limited compared with deep neural networks models in terms of their abilities to model sequence data.

%\subsubsection{Matrix Factorization} Meanwhile, some studies explored the possibility of tailoring the commonly used matrix factorization or metric embedding technique into the next POI recommendation \cite{feng2015personalized,zhao2016stellar,liu2016unified}. For example, Zhao et al. \cite{zhao2016stellar} developed a ranking-based pairwise tensor factorization framework STELLAR to model the interactions among time, user, and POI and provide time-aware successive POI recommendations. Feng et al. \cite{feng2015personalized} presented a personalized ranking metric embedding method (PRME) to jointly model individual preferences and POI transition information. Additionally, a hyperbolic metric embedding model (HME) \cite{feng2020hme} was proposed to learn the representations of check-in data in a low-dimensional hyperbolic space. The main advantage of HME is the hyperbolic space can effectively reserve the underlying hierarchical structures behind check-in records such as the category and region hierarchy. 

% \subsubsection{RNN-based Methods} 
%More recently, researchers turn to neural-based approaches such as RNN, LSTM, or GRU, which can cope with any type of sequential data and be proved effective in many cases. 
More recently, researchers turned to deep learning and advanced embedding methods~\cite{feng2020hme}.
Variants of RNN 
%LSTM/GRU 
were proposed to capture the temporal dynamics and sequential correlations \cite{liu2016predicting,wu2019long,zhao2020go,sun2020go,zhao2020discovering,wu2020personalized}. In 2016, Liu et al. \cite{liu2016predicting} proposed spatial temporal recurrent neural networks (ST-RNN) which incorporated spatio-temporal contexts into RNN layers. In particular, the spatial contexts are represented by geographical distance transition matrices %and temporal features are time intervals.
and time transition matrices encode temporal context. 
% Later on, Yang et al. \cite{yang2017neural} employed RNN and GRU to characterize the sequential relatedness at short-term and long-term sequential contexts. Furthermore, 
LSTM was also adopted to model user's long- and short-term preferences. In LSPL \cite{wu2019long} and PLSPL \cite{wu2020personalized}, the authors trained standard LSTM models for short-term trajectory mining, and general embedding layers to capture users' preference. %Similarly, Sun et al. \cite{sun2020go} proposed a long and short term preference modeling method (LSTPM) which also exploit standard LSTM models to capture the short-term geo-detailed sequence and long-term geo-nonlocal preference. 
%Besides standard LSTM, researchers also developed several variants to cater for next POI recommendation. For instance, 
Zhao et al. \cite{zhao2020go} designed a novel LSTM unit called spatio-temporal gated network (STGN) with two time gates and two distance gates to model the time intervals and distance intervals in both short-term and long-term sequence. %These studies focus on modeling users' long-term preference as well as the short-term trajectory and 
In general, all work above view the next POI recommendation as a sequential prediction task. There are also a few studies that adopted the attention mechanism into this recommendation task, such as DeepMove \cite{feng2018deepmove} and STAN \cite{luo2021stan}. DeepMove \cite{feng2018deepmove} proposed an attention model to capture the multi-level periodicity pattern and a recurrent neural network modeling the sequential transitions for the final recommendation. STAN \cite{luo2021stan} employed the self-attention mechanism to extract the non-adjacent point-to-point interactions where sequential models fail to do.
However, all of these studies overlooking the potential benefits of exploiting generic movements of users. %the common trajectory fragments made by different users.

%Another example is adaptive sequence partitioner (ASP) model \cite{zhao2020discovering}. Compared with LSTM cells, ASP cells have additional cell proposal gate and boundary detector to control the pattern of cell update and capture different granularities in raw input sequence. 

%Additionally, a hyperbolic metric embedding model (HME) \cite{feng2020hme} was proposed to learn the representations of check-in data in a low-dimensional hyperbolic space.

\subsection{Graphs in Location-based Recommendation} 

%\head{Single Graph} 
%Since the intrinsic graph structure in Location-based Social Networks (LBSNs) such as POI geographical graph and user social network, 
In this paper, we aim to take advantage of graph-based methods in next POI recommendation. Graphs-based methods -- such as those that utilise location-based social networks (LBSN) -- provide a powerful paradigm especially for conventional (non-sequential) recommendation tasks.  
%Indeed, a similar theme was adopted by many work in recommendation tasks that involve location-based social networks (LBSN) such as POI geographical graphs and user social networks. 
Indeed,  \cite{yuan2014graph} built a geographical-temporal influences aware graph (GTAG) which is a tripartite graph consisting of POI, session, and user nodes. 
%We can view GTAG graph as an upgraded version of bipartite graph in general online recommenders with additional session nodes. 
However, introducing the session nodes can lead to an explosion of graph size. %Typical LBSN datasets with thousands of POIs and users usually have millions of sessions, which may bring in unnecessary overhead during creating, storing or using GTAG graph.
%
% Each session node corresponding to a visit event created by a user at a specific time. In other words, session nodes have time features, and sit between POI nodes and user nodes. Then, authors developed a relaxed breath-first search algorithm to search on GTAG graph directly and return recommended POIs within at most 6 propagation steps. We can view GTAG graph as an upgraded version of bipartite graph in general online recommenders with additional session nodes. However, introducing the session nodes can lead to the potential overwhelming graph size. Typical LBSN datasets with thousands of POIs and users usually have millions of sessions, which may bring in unnecessary overhead during creating, storing or using GTAG graph.
%
%\head{Multi-Graphs} 
Unlike GTAG, \cite{xie2016learning} constructed four bipartite graphs, including
%for POI recommendation. In particular, four bipartite graphs 
POI-POI, POI-Region, POI-Time and POI-Word interaction graphs, to capture sequential effect, geographical influence, temporal dynamics, and semantic features, respectively. 
%In POI-POI graph, the number of nodes is the twice as many as the number of POIs, and the directed edge connected two successive POIs in sequence. POI-Region graph depicts the geographical information of each POI. POI-Time graph illustrates the check-in frequency of a POI at a time point. The last POI-Word graph links each POI with a set of description words as the POI background information. 
%Equipped with the four graphs, 
The authors then extended LINE network embedding model \cite{tang2015line} to bipartite graphs, and used conditional probability as well as other statistical tools to train graph embeddings. A similar work is GGLR \cite{chang2020learning}, graph-based geographical latent representation model. GGLR utilizes POI-POI graph and POI-User graph. In POI-User bipartite graph, user nodes and POI nodes are connected if the user ever visited the POI in check-in history, which aims to highlight user's preference. 
%\head{Hypergraph} 
%Another important work is LBSN2Vec \cite{yang2019revisiting} where hypergraph is proposed. The hypergraph contains two types of edges: user-user edges and user-time-POI-semantic hyperedges (check-ins). The authors then developed a random-walk method on the hypergraph to learn the node embeddings for downstream tasks such as user-user link prediction and location prediction.

We point out that all studies above are for conventional ({\em not-}``next'') POI recommendation problem. 
%or general LBSNs embedding methods. 
%Graphs in next POI recommendation is less explored as the sequential features usually play a predominant role for next POI prediction. 
To our knowledge, no work exists that explicitly leverages a unified graph structure for next POI recommendation. In a recent work \cite{li2021discovering}, for each POI in a given check-in sequence, the authors randomly sampled previous and next check-ins from other check-in sequences and incorporated these POIs to train an embedding of the current POI. The aim was to capture local (one-hop) transition of POIs and thus the embedding does not explicitly reflect the global (multi-hop) transition patterns. Instead, our \graph is a unified graph structure that manifests global patterns among all POIs. In this sense, {\em our model is the first that utilizes graph-based learning to encode generic transitional information of POIs in a next POI recommendation task}. 

%A recent work used a local-graph liked method to employ collaborative signals \cite{li2021discovering}. Instead of building a normal graph ahead, for each POI in a given sequence, the authors randomly sample the previous check-in and next check-in from other sequences and use these neighbours as the collaborative signals during training. Compared with our methods, we build a global trajectory flow map before training, which is a weighted directed graph and preserves the quantitative local and global collaborative signals, as well as the geographical and meta information of POIs. 

% POIs are nodes that are chronologically ordered
%\todo{Distinguish with "Discovering Collaborative Signals for Next POI Recommendation with Iterative Seq2Graph Augmentation" Global vs local (one-hop)}

\section{Problem Formulation}
Let $U = \{u_1, u_2, \cdots, u_M\}$ be a set of users, $P = \{p_1, p_2, 
\cdots, p_N \}$ be a set of POIs (such as specific restaurants, hotels) and $T=\{t_1,t_2,\cdots,t_K\}$ denotes the set of time stamps, where $M$, $N$, $K$ are positive integers. 
%the total number of users, POIs and time points, respectively. 
Each POI $p\in P$ is denoted by a tuple $p=\langle lat,lon,cat, freq\rangle$ of latitude, longitude, category and check-in frequency, respectively.
We now define several key concepts in the paper. In particular, $cat$ is taken from a fixed list of {\em POI categories} (e.g., ``train station'', ``bar''). % POI features (GPS, category, freq)

% 删除POI的definition，放在上一段中
% \begin{definition}[POI] In LBSNs, a point-of-interest (POI) $p \in P$ is spatial location such as a restaurant, a hotel, or a gym, which associated with a GPS coordinates and other auxiliary information such as category. 
% \end{definition}

\begin{definition}[Check-in] A \emph{check-in} is a tuple $q = \langle u, p, t\rangle\in U\times  P \times T$, indicating that user $u$ visits POI $p$ at  time stamp $t$.
\end{definition}

%\begin{definition}[Check-in sequence] 
All check-in activities created by user $u \in U$ forms a {\em check-in sequence} $Q_{u}= (q_u^1, q_u^2, q_u^3, \cdots )$ where $q_u^i$ is i-th check-in record. Denote the check-in sequences of all users as $Q_U = \{Q_{u_1}, Q_{u_2},\cdots,Q_{u_M} \}$.
%\end{definition}

% From the definition of check-in sequence, we know $Q_{u_i}$ can be arbitrary length and have a wide time range such as a whole year. However, to predict the immediate next POI, the check-ins from few months ago are not critical.

%\begin{definition}[Trajectory] 
As data preprocessing, we split the check-in sequence $Q_{u}$ of any user $u$ into a set of consecutive \emph{trajectories}, namely, $Q_{u}=S^1_{u} \oplus S^2_{u} \oplus \cdots$ where $\oplus$ denotes concatenation. 
%\end{definition}
The length of the trajectories may be different and each of them contains a list of check-ins within a short time interval (e.g., 24 hours).

The goal of {\em next POI recommendation} is to provide a list of possible POIs that a user is inclined to visit next, by learning from the current trajectory and all user's historical check-in records. More formally, given a set of historic trajectories $\{S^i_u\}_{i\in \mathbb{N}, u\in U}$, and a current trajectory $S'=(q_1,q_2,\ldots,q_m)$ of a specific user $u_i\in U$, predict the most likely future POIs $q_{m+1}, q_{m+2},\ldots, q_{m+k}$ that $u_i$ would visit next for a small integer $k\geq 1$ (normally $k=1$).

%The problem of next POI recommendation is thus formulated as given all users' historical check-in records $Q_U$, a recent trajectory $S^{t+1}_{u} \not\subset Q_{u}$, the goal is to predict the most likely next POI for user $u$.

% \begin{definition}[Next POI Recommendation] 
% Given all users' historical check-in records $Q_U$, a recent trajectory $S^{t+1}_{u} \not\subset Q_{u}$, the goal is to predict the most likely next POI for user $u$.
% \end{definition}

\section{Our Approach: \GETNext}
\subsection{Model Structure Overview}
%字体调大，把encoded seq和transformer input连起来，给中间的框加上title (contextual embedding), 右边的框加上title (encoder-decoder?)
\begin{figure*}
	\centering
	\includegraphics[width=0.95\linewidth,height=3.9in]{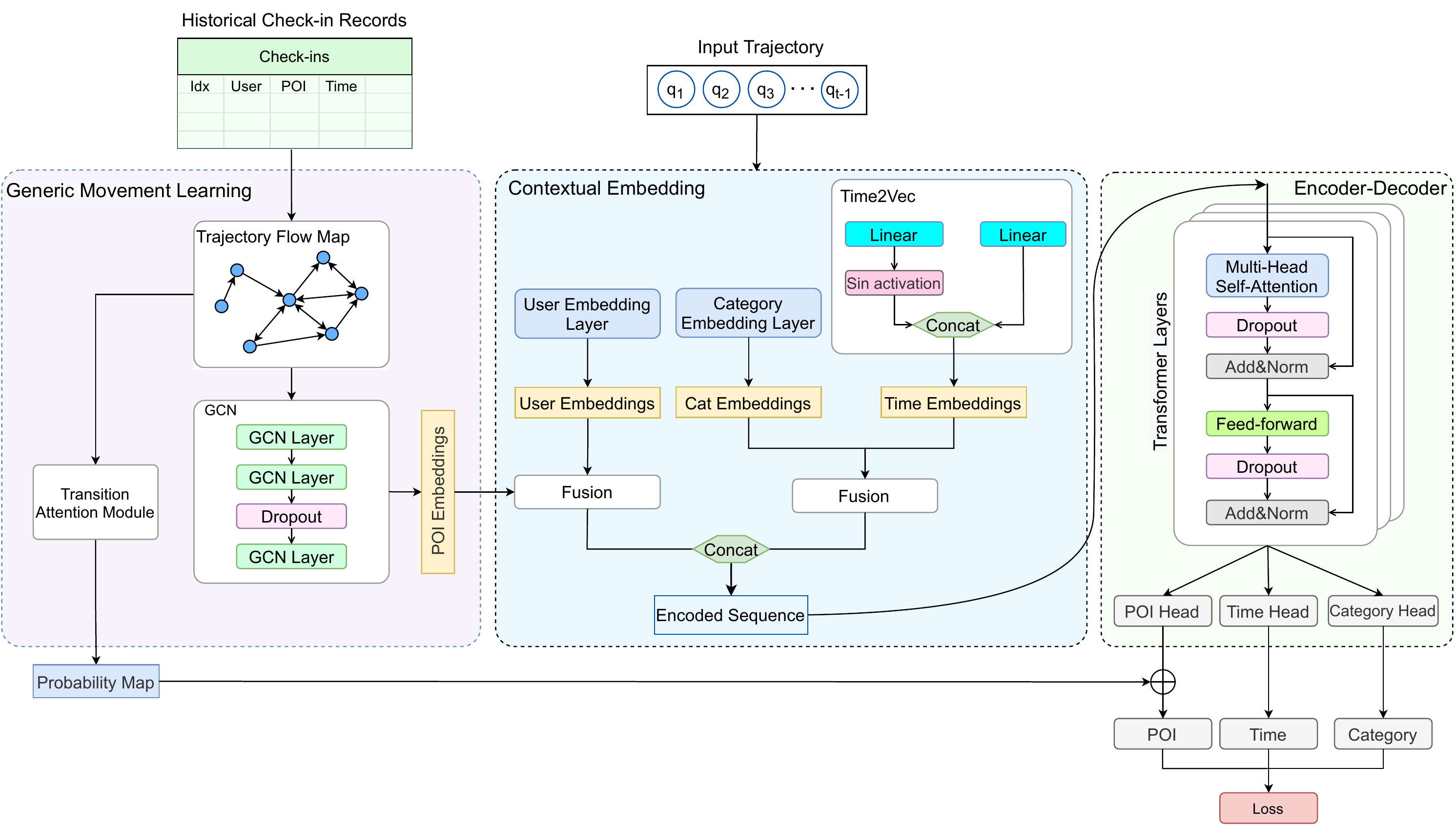}
	\caption{An overview of the proposed GETNext model}
	\label{fig:model-structure}
\end{figure*}

Figure~\ref{fig:model-structure} illustrates the overall design of our proposed \GETNext\;model. The model fuses several key components. First, we define {\em \graph} which summarizes historical trajectories (see Sec.~\ref{sec:graph}). The \graph influences the recommendation output in two ways: 
\begin{enumerate}[leftmargin=*]
\item  A graph neural network (GNN) is trained on  \graph to produce {\em POI embeddings} that encode users' generic movement patterns on each POI, while incorporating the category, location, and check-in frequency of the POI. 
\item  An attention module that takes the adjacency matrix of \graph and node features as input, and produces a {\em transition attention map} that explicitly models transition probabilities between POIs. 
\end{enumerate}
We then define several contextual modules to obtain  {\em user embeddings}, {\em POI category embeddings}, and {\em time encodings} (through a time2vector model) (See Sec.~\ref{sec:context}). 
\begin{itemize}[leftmargin=*]
    \item The user embeddings and the POI embeddings in corresponding trajectories are combined  for better personalization. 
    \item The POI category embeddings and the time encodings are also combined to capture users' temporal preferences of different POI categories (consider, e.g., train stations at peak hours). 
\end{itemize} 
The above allows us to produce a single {\em check-in embedding} vector by unifying the user, POI category, time stamp, and POI information in a check-in record. Each trajectory can then be encoded as a list of such check-in embeddings. We then employ a transformer encoder and multilayer perceptron (MLP) heads to produce a POI prediction. At last, the predicted POI is adjusted by the learned \attnmap with a residual connection. 

%First, we build a \graph from all historical check-in records to leverage the collaborative signals and alleviate the cold-start problem. Then, a graph neural network is trained on the \graph to produce the POI embeddings that incorporate the transition patterns, POI category, geographical location, and POI popularity. Meanwhile, an attention module takes the graph adjacency matrix and node features as the input and output a transition probability map, which influence the final recommended POI directly. Next, user, category embeddings are learned from embedding layers, and a time2vector model to convert time stamps to vectors. After all necessary embeddings obtained, we fuse the user embeddings with POI embeddings in the short trajectory to construct a better personalized context. Moreover, POI category embeddings and time vectors are also blended as we observe the categories usually exhibit strong time preference such as the train station at peak hours. As a result, we convert the user, POI, POI category, and time stamp associated with a check-in record to a single embedding vector. Then each trajectory can be encoded as a list of check-in embeddings and a transformer encoder and multilayer perceptron (MLP) heads are employed to make the prediction of next POI, category and time. As last, we adjust the predicted POI by the learned \attnmap with a residual connection.

\subsection{Learning with Trajectory Flow Map}\label{sec:graph}
\subsubsection{POI Embedding}
We observe that different individuals may share certain  similar trajectory fragments and the same person can repeat a trajectory multiple times. To utilize these common patterns across historical check-in records, we construct an \textit{(user-agnostic) \graph} to provide a global view of users' generic movements among POI. 

\begin{definition}[Trajectory flow map]
Given the set of historic trajectories $\mathcal{S}=\{S^i_u\}_{i\in \mathbb{N}, u\in U}$, a {\em \graph} is an attributed weighted directed graph $\mathcal{G}=(V,E,\ell,w)$ such that
\begin{itemize}[leftmargin=*]%[left margin=*]
    \item the set of nodes is $V=P$ the set of POIs, 
    \item the attribute $\ell(p)$ of each $p\in P$ consists of $(lat,lon,category, freq)$ where $(lat,lon)$ is the coordinate of $p$, $category$ is the category of $p$, and $freq$ is the number of times $p$ occurs in trajectories in $\mathcal{S}$.  
    \item there is an edge from $p_1$ to $p_2$ if $(p_1,p_2)$ appears in a trajectory $S^i_u$ in $\mathcal{S}$, i.e., they are visited consecutively.
    \item the weight $w(p_1,p_2)$ of any edge $(p_1,p_2)$ equals the number of times $(p_1,p_2)$ appears in any trajectory in $\mathcal{S}$. 
\end{itemize}
\end{definition}

We remark that apart of indicating the movements of users among POIs, the \graph also implicitly reveals certain spatial proximity between POIs. The intuition is that if two POIs are far from each other, they are less likely to be checked in successively.

%\subsubsection{Graph Convolution Network}
Given the \graph $\mathcal{G}$, our next step is to learn a vectorized representation of POIs that encodes the common POI transition patterns and the attributes of POIs. For this we utilize  graph convolution network (GCN).
In order to take full advantage of the topological information of $\mathcal{G}$, we use the spectral GCN \cite{kipf2016semi}. In particular, let $\mathbf{A} \in \mathbb{R}^{N\times N}$ denote the adjacency matrix of  $\mathcal{G}$, we first compute the normalized Laplacian matrix as
\begin{equation}
\mathbf{\Tilde{L}} = (\mathbf{D}+\mathbf{I}_N)^{-1}(\mathbf{A}+\mathbf{I}_N)
\end{equation}
where $\mathbf{D}$ is the degree matrix and $\mathbf{I}_N$ is the identity matrix of $\mathcal{G}$. Next, 
let $\mathbf{H}^{(0)} = \mathbf{X}\in \mathbb{R}^{N\times C}$ be the input node feature matrix. We define the propagation rule between GCN layers as
\begin{equation}
\mathbf{H}^{(l)} = \sigma\left(\mathbf{\Tilde{L}} \mathbf{H}^{(l-1)} \mathbf{W}^{(l)} + b^{(l)}\right)
\end{equation}
where $\mathbf{H}^{(l-1)}$ denotes the input signals of the $l$-th layer for any $l > 0$, $\mathbf{W}^{(l)} \in \mathbb{R}^{C\times \Omega}$ represents the model weights matrix at the $l$-th layer, the corresponding bias $b^{(l)}\in \mathbb{R}^{C\times \Omega}$, and $\sigma$ is a leaky ReLU activation function  for non-linearity (with leaky rate 0.2). %The output of $l$-th layer becomes the input of $(l+1)$-th layer with shape $\mathbf{H}^{(l+1)}\in \mathbb{R}^{N\times \Omega}$. 

%In the GCN layer, we first project the node features to a high dimensional space by weights matrix $\mathbf{W}$ and then inject the transition probability information by multiply with the normalized laplacian matrix $\mathbf{\Tilde{L}}$. At last, a bias term is added. % as the standard convolutional kernel usually does. 
From a spatial perspective, at each iteration, the GCN layer updates a node's embedding by aggregating its neighborhood information together with the node's own embedding.  
We stack $l^*$ GCN layers to increase the model's expressiveness. Dropout is employed before the last layer. The output of the GCN module can be written by:
\begin{equation}
\mathbf{e}_{\mathrm{P}} = \mathbf{\Tilde{L}}\; \mathbf{H}^{(l^*)}\mathbf{W}^{(l^*+1)}+b^{(l^*+1)} \in \mathbb{R}^{N\times \Omega}
\end{equation}
% \begin{equation}
% \mathbf{e}_{P} = \sigma\left(\mathbf{\Tilde{L}}\; \sigma\left(\mathbf{\Tilde{L}} \mathbf{H}^{(l-1)} \mathbf{W}^{(l-1)} + b^{(l-1)}\right) \mathbf{W}^{(l)}+b^{(l)}\right) \in \mathbb{R}^{N\times \Omega}.
% \end{equation}
Finally, the embedding $\mathbf{e}_{p_i}$ of  POI $p_i$ is the $i$-th row of the $N\times \Omega$ matrix $\mathbf{e}_\mathrm{P}$. 
%
%The output embedding matrix $\mathbf{e}_{P} \in \mathbb{R}^{N\times \Omega}$ is the vectorized representation of all $N$ POIs. 
%
Loosely speaking, the embedding of a POI $p$ indicates the position of $p$ within the historical trajectories of {\em all} users and thus captures generic movement patterns at $p$. It will be in turn fed to the transformer downstream to model users' visiting behaviors. Note that even when the current trajectory is short, the POI embeddings nevertheless provides rich information to the prediction model.
%We set the POI embedding size $\Omega$ to 128 in experiments.
% In experiments, we build the model with 3 GCN layers. 

\subsubsection{Transition Attention Map}\label{sec:attn_map} The POI embeddings learned from the graph $\mathcal{G}$ capture generic movement patterns only \emph{implicitly}. To amplify the impact of the collective signals, we propose a novel {\em transition attention map} to \emph{explicitly} model the transition probabilities from one POI to another. As stated above, these transition probabilities will be used to adjust the final prediction.% (See Sec.~\ref{}).
Given the input node features and $\mathcal{G}$, we compute the attention map $\mathbf{\Phi}$ as:
\begin{equation}
\mathbf{\Phi}_1 = (\mathbf{X} \times \mathbf{W}_1)\times \mathbf{a}_1 \in \mathbb{R}^{N \times 1}
\end{equation}
\begin{equation}
\mathbf{\Phi}_2 = (\mathbf{X} \times \mathbf{W}_2)\times \mathbf{a}_2 \in \mathbb{R}^{N \times 1}
\end{equation}
\begin{equation}
\mathbf{\Phi} = (\mathbf{\Phi}_1 \times \mathbf{1}^\top + \mathbf{1} \times \mathbf{\Phi}_2^\top) \odot (\mathbf{\Tilde{L}}+J_N) \in \mathbb{R}^{N \times N}
\end{equation}
where %$\mathbf{X} \in \mathbb{R}^{N\times C}$ is node feature matrix;
$\mathbf{W}_1$, $\mathbf{W}_2 \in \mathbb{R}^{C \times h}$ are two trainable feature transformation matrices; $\mathbf{a}_1$, $\mathbf{a}_2 \in \mathbb{R}^{h}$ are two learnable vectors used to construct an $N\times N$ attention matrix by the broadcast add operation; $\mathbf{1}$ is an all-ones vector with shape $\mathbb{R}^{N \times 1}$;
$J_N$ is the matrix of ones, and $\odot$ stands for element-wise multiplication. We shift the range of the normalized Laplacian matrix $\mathbf{\Tilde{L}}$ from $[0,1]$ to $[1,2]$ to avoid zero values.

The $i$-th row of the transition attention map $\mathbf{\Phi}$ indicates the (unnormalized) probability of moving to each POI from the POI $p_i$. Given the last POI in the current trajectory, we lookup the transition probabilities stored in the corresponding row of $\mathbf{\Phi}$, and use these probabilities to adjust recommendation results produced by the later transformer module. %By doing so, we try to make the collaborative signals contributing to the final prediction more significant and explicitly. The experiment results show that with this short-cut attention module, the performance can be further improved.

% rather than the graph attention network (GAT) because we would like to leverage the weighted adjacency matrix explicitly during the neighborhood aggregation phase in GCN. But GAT %GAT只把adj mtx作为mask，没有很好的利用到从all historical data中得到的freq信息.

\subsection{Contextual Embedding Module}\label{sec:context}
Spatio-temporal contexts and user preferences are key factors for personalized next POI recommendations~\cite{feng2020hme,zhao2020go,liu2016predicting}. We proceed to present our contextual embedding module for fusing context information including user embeddings, POI embeddings, POI category embeddings and time encoding. 

% \subsubsection{User Embeddings}
% To capture the user's general taste, we train an embedding layer that project each user to a low-dimensional vector. The embedding of each user is learned from his/her historical check-in sequences.

% In particular, take user $u$ as an example. The corresponding check-ins sequences is $Q_u$ and a single check-in records is $q_u^i=(u,p_i,t_i)$. Though POI $p_i$ and time point $t_i$ change each time, but the user remain consistent. Therefore, by training an embedding layer for user $u$, we try to encode the general tastes of $u$ into the user embedding $\mathbf{e}_{u} \in \mathbb{R}^{\Omega}$ with the same embedding size of POI. Denoted the embedding layer as $f_\text{embed}(\cdot)$, the learned user embedding can be written by
% \begin{equation}
% \mathbf{e}_{u} = f_\text{embed}(u) \in \mathbb{R}^{\Omega}.
% \end{equation}

\subsubsection{POI-User Embeddings Fusion}
The POI embeddings are learned from the trajectory flow map and omit user-specific patterns. To capture a specific user $u$'s general behaviors, we train an embedding layer that project each user to a low-dimensional vector. The embedding of each user is learned from his/her historical check-in sequences. Formally, the {\em user embedding} of $u$ is
\begin{equation}
\mathbf{e}_{u} = f_\text{embed}(u) \in \mathbb{R}^{\Omega}.
\end{equation}
%where $\Omega$ is the same dimensions as POI embeddings. 

In order to construct the representation of each check-in activity, a straightforward solution is to concatenate the POI embedding and user embedding. We feed the concatenated vector into a dense layer to fine-tune the fused embedding and increase its representational power. The output can be denoted as
\begin{equation}
\mathbf{e}_{p,u}= \sigma(\mathbf{w}_{p,u}[\mathbf{e}_{p};\mathbf{e}_{u}]+b_{p,u}) \in \mathbb{R}^{\Omega\times 2},
\end{equation}
where $\mathbf{w}_{p,u}$ and $b_{p,u}$ are weights vector and the bias, respectively, and $[\cdot;\cdot]$ represents the concatenation. The dimension of the output embedding is twice as large as the POI embedding or user embedding. In other words, the size of embedding vector remains unchanged after the fusion.

\subsubsection{Time-Category Embeddings Fusion}

\begin{figure}
	\centering
	\begin{subfigure}[b]{0.48\linewidth}
		\centering
		\includegraphics[width=\linewidth]{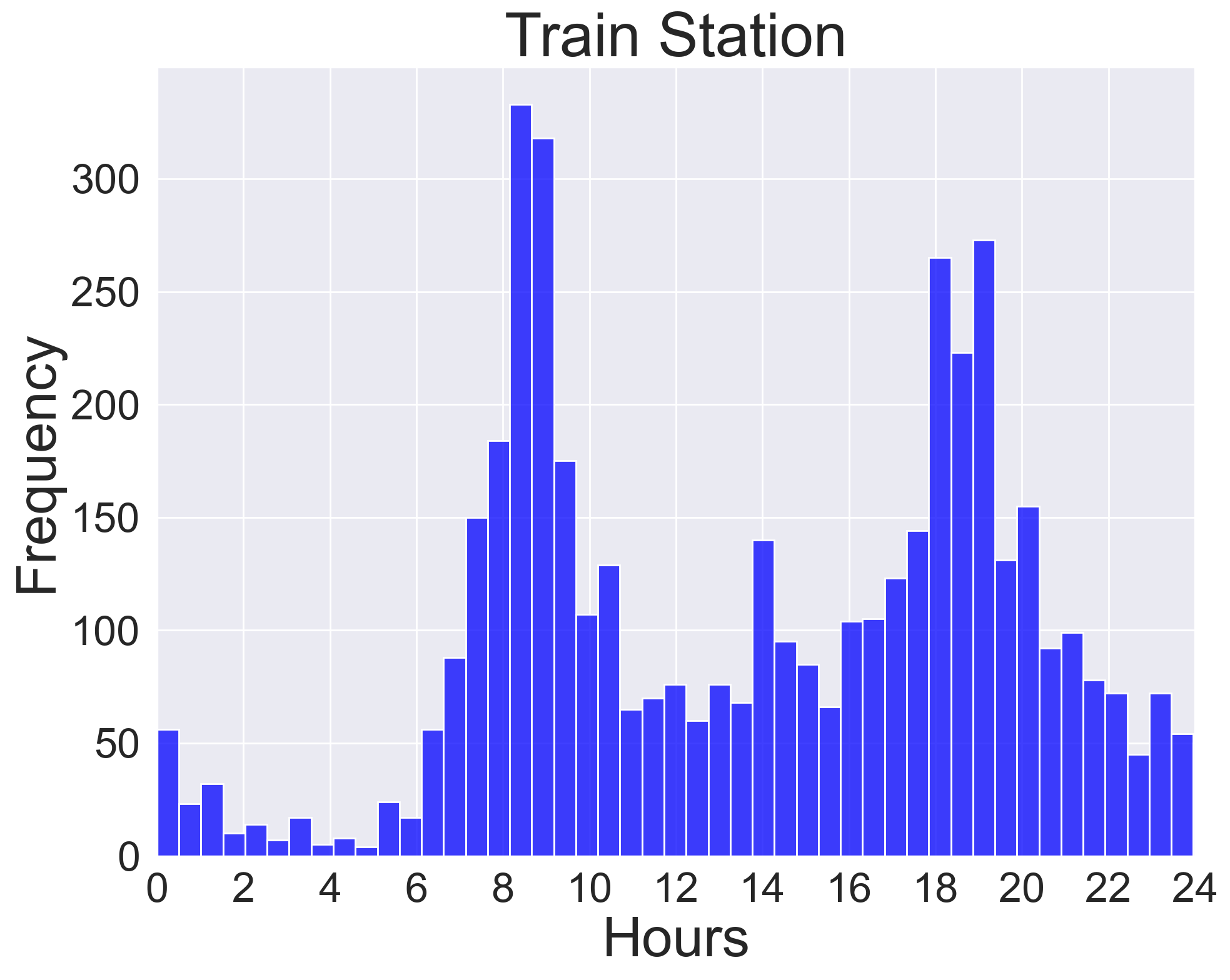}
		\caption{``Train station''}
	\end{subfigure}
	\hfill
	\begin{subfigure}[b]{0.48\linewidth}
		\centering
		\includegraphics[width=\linewidth]{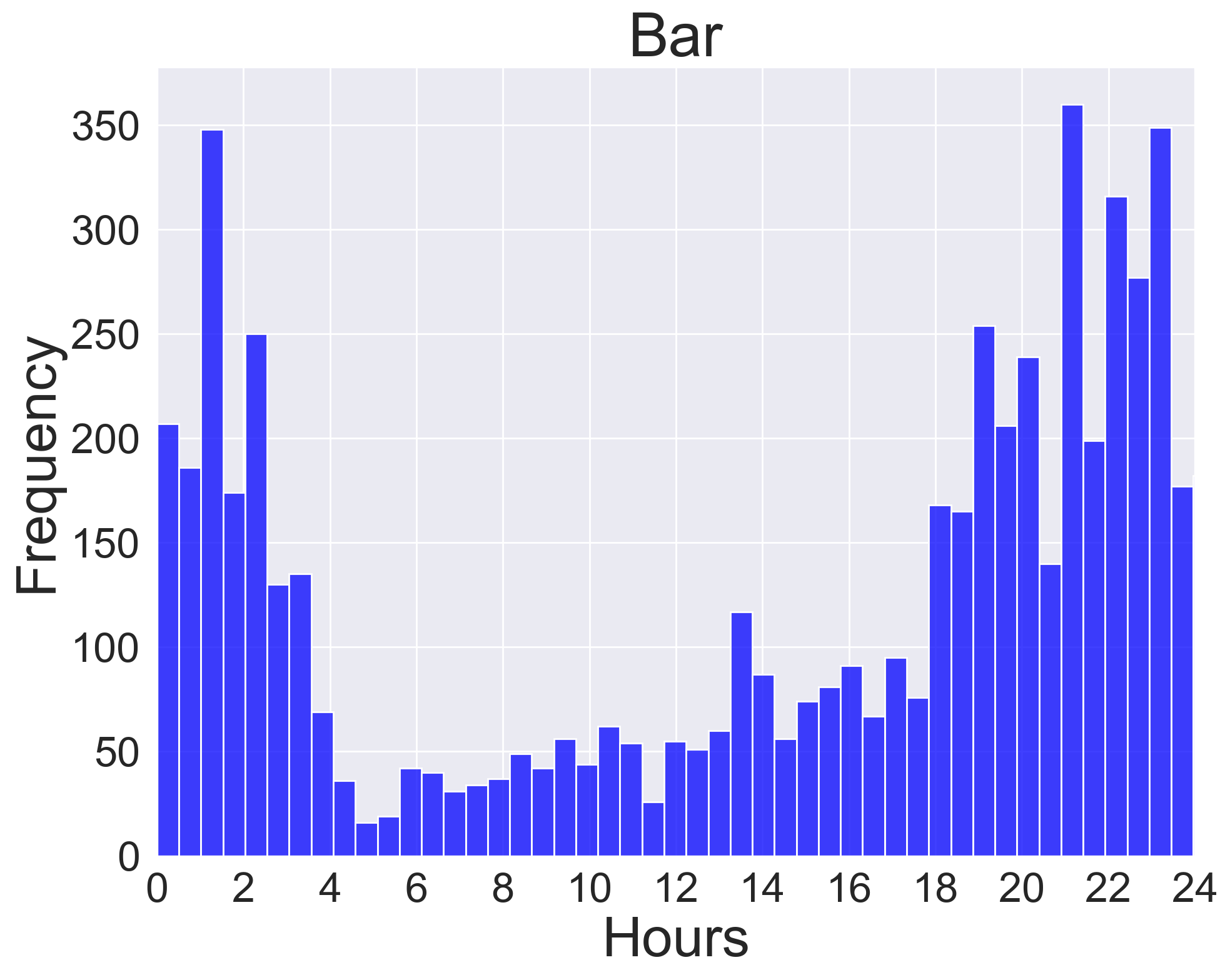}
		\caption{``Bar''}
	\end{subfigure}
	\caption{Average hourly check-in frequency of two POI categories (``train station'' and ``bar'') in NYC dataset}
	\label{fig:time-dst}
	%\vspace{-0.2cm}
\end{figure}

%\subsubsection{Time2Vector}
The visiting behaviors of users are naturally time-dependent. For example, Fig.~\ref{fig:time-dst} shows the check-in frequencies of two POI categories in a day. ``Train station'' has two clear peaks during the rush hours (around 8AM and 6PM). On the contrary, ``bars'' shows a diametrically opposite pattern where most check-in activities happen after 6PM. 
Such observations motivate us to consider temporal patterns of POI categories for next POI recommendation. For example, a train station -- rather than a bar -- should be recommended at 8AM.

It is worth noting that check-ins at individual POIs may also exhibit certain temporal patterns. However, because of scarcity of check-ins and noise in data, the temporal patterns of individual POIs are far less clear and stable as those of categories. Take NYC dataset as an example, which contains 227k check-in records, 38k POIs, and 400 categories. On average, each POI only gets less than 6 check-ins. %Not even enough to cover the 24 hours in a day. 
The check-ins at the category level, in contrast, is more than 570 per category. We thus explore the temporal patterns of POI categories instead of individual POIs.

%even without any context, a sensible recommender should output the train station with higher probability over the bar at 8AM. 

Similar to the POI-User embedding fusion module, we first encode POI category and time. 
%A primitive solution of encoding the time information is to split a day into 24 hour slots and represent each time slot by one-hot encoding. However, this solution is rather rigid and lack of representation power to handle a great number of POI categories with various patterns. 
We adopt  time2vector, a state-of-the-art time encoding model~\cite{kazemi2019time2vec}.
Specifically, we divide the 24 hours in a day to 48 slots with 30 minutes per slot. We project a (scalar) time value to one of the time slots. The embedding of a time slot $t$, denoted as $\mathbf{e}_{t}$, is a vector of length $k+1$. The $i$-th element is defined as:
\begin{equation}
\mathbf{e}_{t}[i]= \begin{cases}\omega_{i} t+\varphi_{i}, & \text { if } i=0 . \\ \sin\left(\omega_{i} t+\varphi_{i}\right), & \text { if } 1 \leq i \leq k .\end{cases}
\end{equation}
where $\omega$ and $\varphi$ are learnable parameters. The $\sin$ activation function is used to capture periodic patterns.

% \todo{Correlation between single POI and time is not as clear as category and time; do some analysis to support argument}

%Like user embedding, we 
We employ another embedding layer for POI categories. The embedding of a category $c$ is  
\begin{equation}
\mathbf{e}_{c} = f_\text{embed}(c) \in \mathbb{R}^{\Psi},
\end{equation}
where $\Psi$ is the embedding dimension.

Next, we fuse the time embedding $\mathbf{e}_{t}$ and category embedding $\mathbf{e}_{c}$ to form a single category-time representation by a dense layer:
\begin{equation}%\operatorname{Leaky\_ReLU}\sigma
\mathbf{e}_{c,t}= \sigma(\mathbf{w}_{c,t}[\mathbf{e}_{t};\mathbf{e}_{c}]+b_{c,t}) \in \mathbb{R}^{\Psi\times 2},
\end{equation}
where $\mathbf{w}_{c,t}$ is a learnable weight vector and $b_{c,t}$ is the bias. 
%In other words, for each check-in, the corresponding time embedding and POI category embedding are concatenated directly. Then, 
%We then apply a dense layer to handle the correlations between time and category to increase the representation power of the fused embeddings.

Finally, the embedding of a check-in $q=\langle p,u,t\rangle$ where POI $p$ has category $c$ is the concatenation $\mathbf{e}_{q} = [\mathbf{e}_{p,u}; \mathbf{e}_{c,t}]$. As a result, each input trajectory $(q_1,\ldots,q_s)$ is represented by a list of check-in embeddings $(\mathbf{e}_{q_1},\ldots,\mathbf{e}_{q_s})$. We will feed the encoded sequence to the transformer encoder.

\subsection{Transformer Encoder and MLP Decoders}\label{sec:transformer}
\subsubsection{Transformer Encoder}
The transformer is a purely attention based encoder-decoder network, which was firstly proposed by Vaswani et al \cite{vaswani2017attention} in 2017. Due to its distinguished computational efficiency and outstanding performance compared with traditional models, transformers has become the paradigm-of-choice in NLP, and more recently in image recognition \cite{dosovitskiy2020image}, video understanding \cite{neimark2021video}, time series analysis \cite{lim2021temporal}.

%are taking the natural language (NLP) processing world by storm. Recently, the transformers are start to roaring outside the NLP filed and many researchers investigated the probability of using transformer in other tasks such as image recognition \cite{dosovitskiy2020image}, video understanding \cite{neimark2021video}, time series analysis \cite{lim2021temporal}, etc.

We adopt transformer for next-POI recommendation as it is a natural sequence prediction task. 
%can be regarded as a sequence prediction problem inherently. % Thus, using transformer to predict the next move is a reasonable choice. 
Since the goal is to predict only the next immediate POI given a check-in sequence, 
%the transformer decoder, which decodes the whole sequence, is not necessary. We 
we adopt the transformer encoder only followed by several MLP heads (without the transformer decoder). For the encoder, we stack several standard transformer encoder layers with positional encoding. Each layer consists of a multi-head self-attention module followed by a position-wise fully-connected network. Residual connections and normalization are applied to both modules. 

Formally, given an input trajectory $S_u = (q_u^1, q_u^2, \dots, q_u^k)$, to predict the POI in the next check-in activity $q_u^{k+1}$, we firstly take the check-in embedding of each historical check-in $q_u^i$, $1\leq i\leq k$, as defined in Sec~\ref{sec:context}. We then stack these check-in embeddings to form an input tensor of the first encoder layer, which can be denoted as $\mathcal{X}^{[0]} \in \mathbb{R}^{k\times d}$, where $d$ represents the embedding dimension of each check-in record. Particularly, $d$ equals to the length of $\mathbf{e}_{q}$, i.e., $d=2\times (\Omega + \Psi)$. It is worth noting that the embedding length $d$ remains unchanged across all encoder layers to facilitate residual connections and layers stacking. In other words, ignoring padding, the output shape of the encoder layer $l$ coincide with the input shape, i.e., $\mathcal{X}^{[l+1]} \in \mathbb{R}^{k\times d}$. 

For layer $l$, the input $\mathcal{X}^{[l]}\in \mathbb{R}^{k\times d}$ is firstly transformed by a multi-head self-attention module (the number of heads is 2 in our experiments). For the first attention head, the output is:
\begin{equation}
\boldsymbol{S} = \mathcal{X}^{[l]}\boldsymbol{W}_{q}  (\mathcal{X}^{[l]}\boldsymbol{W}_{k} )^T \in \mathbb{R}^{k \times k}
\end{equation}
\begin{equation}
\boldsymbol{S}^{\prime}_{i,j}=\frac{\exp \left(\boldsymbol{S}_{i, j} %(i,j) 
\right)}{\sum_{j=1}^{d} \exp \left(\boldsymbol{S}_{i, j} %(i,j)
\right)} 
\end{equation}
\begin{equation}
\operatorname{head}_{1} =  \boldsymbol{S}^{{\prime}}\mathcal{X}^{[l]}\boldsymbol{W}_{v}  \in \mathbb{R}^{k \times d/h},
\end{equation}
where $\boldsymbol{W}_{q} \in \mathbb{R}^{d \times d/h}$,  $\boldsymbol{W}_{k} \in \mathbb{R}^{d \times d/h}$ and $\boldsymbol{W}_{v} \in \mathbb{R}^{d \times d/h}$ are learnable weight matrices corresponding to ``query'', ``key'' and ``value'', respectively. Here we ignore the bias in equations. The dot product attention, i.e., $\boldsymbol{S}_{i, j}$ indicates the correlations between the $i$-th and $j$-th check-in activities. Next, softmax function is applied to assure the attention weights sum to one. After projecting the input data to an output feature space by $\boldsymbol{W}_{v}$, we use the learned attention matrix to adjust the contribution of each check-in record. Last, we stack different attention heads and employ another linear transformation $\boldsymbol{W}_{o} \in \mathbb{R}^{d \times d}$ to merge the representations from different attention spaces:
\begin{equation}
\operatorname{Multihead}(\mathcal{X}^{[l]}) = [\operatorname{head}_{1};\cdots;\operatorname{head}_{h}]\times \boldsymbol{W}_o \in \mathbb{R}^{k\times d}.
\end{equation}

Furthermore, the layer norm and residual connection is applied to the attention module. Therefore, the final output of the attention module can be written as
\begin{equation}
\mathcal{X}^{[l]}_\text{attn} = \operatorname{LayerNorm}\left(\mathcal{X}^{[l]} + \operatorname{Multihead}(\mathcal{X}^{[l]})\right).
\end{equation}

In each encoder layer, a fully-connected (FC) network is attached after the attention model. Denote the output of multihead attention module as $\mathcal{X}^{[l]}_\text{attn}$. The FC network can be represented as 
\begin{equation}
\mathcal{X}^{[l]}_\text{FC} = \operatorname{ReLU}(\mathbf{W}_1\mathcal{X}^{[l]}_\text{attn}+b_1)\mathbf{W}_2+b_2
\in \mathbb{R}^{k\times d},
\end{equation}
where $\mathbf{W}_1$, $\mathbf{W}_2$ are trainable weight matrices and $b_1$, $b_2$ are biases. Similarly, the output of the $l$-th encoder layer is
\begin{equation}
\mathcal{X}^{[l+1]} = \operatorname{LayerNorm}(\mathcal{X}^{[l]}_\text{attn} + \mathcal{X}^{[l]}_\text{FC}) \in \mathbb{R}^{k\times d}.
\end{equation}

%\subsection{Decoders and Loss}
\subsubsection{MLP Decoders}
The transformer encoder layers distill the useful information from the input trajectory check-in embeddings to a feature space. In order to predict the user's next move, we replace the transformer decoder with several multi-layer perceptron (MLP) decoders. In particular, we employ three MLP heads to predict the next POI, visit time, and POI category, respectively. Denote the  output of the encoder as $\mathcal{X}^{[l^*]}$, the MLP heads can be written by

\begin{align}
    \mathbf{\hat{Y}}_\text{poi} = \mathcal{X}^{[l^*]}\mathbf{W}_\text{poi} + b_\text{poi} \\
    \mathbf{\hat{Y}}_\text{time} = \mathcal{X}^{[l^*]}\mathbf{W}_\text{time} + b_\text{time} \\
    \mathbf{\hat{Y}}_\text{cat} = \mathcal{X}^{[l^*]}\mathbf{W}_\text{cat} + b_\text{cat}
\end{align}
where $\mathbf{W}_\text{poi} \in \mathbb{R}^{d\times N}$, $\mathbf{W}_\text{time} \in \mathbb{R}^{d\times 1}$, $\mathbf{W}_\text{cat} \in \mathbb{R}^{d\times \Gamma}$ are weights in MLP, and $\Gamma$ represents the number of POI categories. For the output of POI head $\mathbf{\hat{Y}}_\text{poi} \in \mathbb{R}^{k\times N}$, we only concern about the last row which corresponds to the POI recommendation for the future move. We in addition combine this POI recommendation with the transition attention map defined in Sec.~\ref{sec:attn_map}. The final recommendation is
\begin{equation}
    \mathbf{\hat{y}}_\text{poi} = \mathbf{\hat{Y}}_\text{poi}^{(k\cdot)}+\Phi^{(p_k\cdot)} \in \mathbb{R}^{1\times N},
\end{equation}
where $\mathbf{\hat{Y}}_\text{poi}^{(k\cdot)}$ represents the $k$-th row of $\mathbf{\hat{Y}}_{poi}$ and $\Phi^{(p_k\cdot)}$ is the $p_k$-th row of transition map $\Phi$, $p_k$ stands for the POI of check-in $q_u^k$.

As shown above, besides the POI head, we also add the time head and category head. The main reason we predict the next check-in time along with POI is the following: The time gaps between two check-ins fluctuate considerably (e.g., between half an hour to as long as 6 hours). This is reasonable as users spend unequal lengths of time in different POIs or forget to record certain check-ins. However, such fluctuation has a considerable impact to prediction. Indeed, a user should receive different recommendations at 5PM for the next hour and for the next 5 hours.  
%Ignoring the possible future temporal context would jeopardize the recommendation performance. Therefore, 
We therefore recommend the next POI as well as the next check-in time, and use the time head as a calibration of the time modeling.
Moreover, the category head is employed to regulate next POI prediction as forecasting the next POI category is easier than exact POI prediction. 
%However, we did not just select the POI from the predict category but use two parallel heads. The main aim is to use the predicted category information as the supplementary instead of the predominant factor.

% \todo{write more about time and cat decoders; why we need three heads instead of only POI head}

\subsubsection{Loss}
%Although we only measure the performance of next POI recommendation, but 
All MLP heads in the decoder are taken into the consideration for training where outputs from the time and category heads act as regularization terms. In other words, we calculate a weighted sum of the losses of all MLP heads. Cross entropy is used as the loss function for POI and POI category prediction. The mean squared error (MSE) is used for the performance of time prediction. Moreover, since we normalized the in-day 24 hours to [0, 1], the scale of time loss is significantly smaller than the other two. To balance the magnitude of gradients with other losses, the time loss term is amplified 10-fold, i.e., %we set scaling coefficient $\alpha = 10$. T
the final loss is
\begin{equation}
    \mathcal{L}_\text{final} = \mathcal{L}_\text{poi}+10\times \mathcal{L}_\text{time}+\mathcal{L}_\text{cat}.
\end{equation}

\section{Experiments}\label{sec:experiment}
In this section, we evaluate our proposed model on real-world datasets.

\subsection{Experimental Setup}
\subsubsection{Datasets}
We conduct experiments on three public datasets collected from location-based service platforms: FourSquare-NYC \cite{yang2014modeling}, FourSquare-TKY \cite{yang2014modeling}, and Gowalla-CA \cite{yuan2013time}. FourSquare-NYC was collected from Apr. 2012 to Feb. 2013 in New York City, and FourSquare-TKY  from Tokyo during the same time period. %Both datasets were collected from FourSquare platform.
%and provided by Yang et al. \cite{yang2014modeling}. 
Gowalla-CA consists of check-ins between Feb. 2009 and Oct. 2010 within California and Nevada from Gowalla \cite{yuan2013time}. Each record contains user, POI, POI category, GPS coordinates, and timestamp. 
%
%\subsubsection{Preprocessing} 
For all three datasets, we exclude unpopular POIs that have less than 10 check-in records, and also filter out users with fewer than 10 check-in history. Next, users' entire check-in sequence were broken into trajectories with 24-hour intervals. In rare cases, the trajectory contains only a single  check-in; these cases are eliminated from the dataset. Next we split the dataset into train/validation/test sets in chronological order. The first 80\% check-ins are the training set and used to build the trajectory flow map $\mathcal{G}$, the middle 10\% are validation set and the remaining 10\% form the test set. Had user or POI not appeared in training but emerged in test, we skip that user or POI when measuring the prediction performance. 
Key statistics of datasets are shown in Table~\ref{tbl:dataset_statistics}.

\begin{table}[!htbp]
%\vspace{-0.2cm}
\caption{Dataset statistics}
\label{tbl:dataset_statistics}
\begin{center}
\begin{tabular}{llllll}
\toprule  & \#user & \#poi & \#cat& \#checkin & \#trajectory \\
\hline NYC & 1,075 & 5,099 & 318 & 104,074 & 14,160 \\
TKY & 2,281 & 7,844 & 291 & 361,430 & 44,692 \\
CA & 4,318 & 9,923 & 301 & 250,780 & 32,920 \\
\bottomrule
\end{tabular}
\end{center}
%\vspace{-0.3cm}
\end{table}

\subsubsection{Evaluation metrics}
We compute the accuracy@$k$ (Acc@$k$) and mean reciprocal rank (MRR), which are common metrics in recommender systems. Accuracy@$k$ indicates whether the true POI appears in the top-$k$ recommended POIs. As acc@$k$ views the top-$k$ recommendations as an unordered list while ignoring the ordering of the correct prediction, we employed  MRR which measures the index of the correctly recommended POI in the ordered result list. 
Given a dataset with $m$ samples (trajectories), define
$$
\text{Acc}@k=\frac{1}{m}\sum_{i=1}^{m} \mathbbm{1}(rank \leq k)
$$
% $$
% \text{MAP}@k=\frac{1}{m}\sum_{i=1}^{m} \frac{\mathbbm{1}(rank \leq k)}{rank}
% $$
$$
\text{MRR}=\frac{1}{m}\sum_{i=1}^{m}\frac{1}{rank}
$$
where $\mathbbm{1}$ is the indicator function. It returns 1 if the condition is true, otherwise 0. Rank represents the rank of the true next POI in the recommended ordered list. In general, for all these metrics, the larger the value, the better the performance.

\subsubsection{Baselines}
We adopt the following baselines.
\begin{itemize}[leftmargin=*]
    \item MF \cite{koren2009matrix} is a classical methods in many recommendation problems. It learned the latent representation of users and POIs by Matrix Factorization.
    \item FPMC \cite{rendle2010factorizing} combined Matrix Factorization and Markov Chain together to model both user long-term preference and sequential behavior.
    \item LSTM \cite{hochreiter1997long} is a variant of RNN model to handle sequential data. Compared with standard RNN model, LSTM models both short-term and long-term sequential patterns.
    \item PRME \cite{feng2015personalized} proposed a pair-wise embedding method named personalized ranking metric embedding to capture the user preference and sequential transition between POIs.
    \item ST-RNN \cite{liu2016predicting} adopted the time, distance transition matrix to model the local temporal and spatial contexts in additional to a RNN for user's sequential patterns capturing.
    \item STGN \cite{zhao2020go} extends the conventional LSTM by adding spatial gates and temporal gates to capture user's preference in space and time dimension. %Spatio-Temporal Gated Network 
    \item STGCN \cite{zhao2020go} is an updated version of STGN which used coupled input and forget gates.
    %\item PLSPL \cite{wu2020personalized} combined user's long term and short term preference by personalized linear layers for different users. The long term taste learned by attention mechanism and short term preference modeled by LSTM.
    \item PLSPL \cite{wu2020personalized} learned user's long term taste by attention mechanism and short term preference with LSTM, and combined them by personalized linear layers.
    \item STAN \cite{luo2021stan} utilizes the spatiotemporal information of checkins along the trajectory with self-attention layers to capture the point-to-point interaction between non-adjacent check-ins.
\end{itemize}

\subsubsection{Experiment Settings}
We developed our model using the PyTorch framework and conducted experiments on the following hardware platform (CPU: AMD Ryzen 9 5900X, GPU: NVIDIA GeForce RTX 3090). The key hyper-parameter settings in our model are listed below. The embedding dimensions of POI and user are both $\Omega=128$. The  time, POI category embedding length are $\Psi=32$. The GCN model has three hidden layers with 32, 64, 128 channels each. The transition attention module convert the input node features to the 128-dim vector. For transformer, we stacked two encoder layers. The dimensions of the feed-forward network in the transformer encoder layer is 1024, and two attention heads are used in the multi-headed attention module. Moreover, we employed the Adam optimizer with 1e-3 learning rate and 5e-4 weight decay rate. Dropout are enabled in both GCN model and Transformer encoder with rate 0.3. Another important parameter is the weight of time loss $\alpha$ where we set to 10 to match the scale of POI loss and category loss. We use the same settings in three datasets and run each model 200 epochs with batch size 20. %Last, random seed of Python, Numpy and PyTorch are set to be 42 for reproducibility purpose. \todo{list all hyper-parameters value}

\begin{table*}[htbp]
\caption{Performance comparison in Acc@k and MRR on three datasets}
\label{tbl:result}
\small
\setlength{\tabcolsep}{0.4em} % for the horizontal padding
\def\arraystretch{1.2} % for the vertical padding
\begin{tabular}{llllll|lllll|lllll}
\toprule
& \multicolumn{5}{c}{NYC}  & \multicolumn{5}{c}{TKY} & \multicolumn{5}{c}{CA} \\
\cmidrule(lr){1-6}
\cmidrule(lr){7-11}
\cmidrule(lr){12-16}
& Acc@1 & Acc@5 & Acc@10 & Acc@20 & MRR 
& Acc@1 & Acc@5 & Acc@10 & Acc@20 & MRR
& Acc@1 & Acc@5 & Acc@10 & Acc@20 & MRR   \\
\cmidrule(lr){1-6}
\cmidrule(lr){7-11}
\cmidrule(lr){12-16}
 
MF 
& 0.0368 & 0.0961 & 0.1522 & 0.2375 & 0.0672 
& 0.0241 & 0.0701 & 0.1267 & 0.1845 & 0.4861 
& 0.0110 & 0.0442 & 0.0723 & 0.1190 & 0.0342  \\

FPMC 
& 0.1003 & 0.2126 & 0.2970 & 0.3323 & 0.1701
& 0.0814 & 0.2045 & 0.2746 & 0.3450 & 0.1344 
& 0.0383 & 0.0702 & 0.1159 & 0.1682 & 0.0911  \\

LSTM 
& 0.1305 & 0.2719 & 0.3283 & 0.3568 & 0.1857 
& 0.1335 & 0.2728 & 0.3277 & 0.3598 & 0.1834 
& 0.0665 & 0.1306 & 0.1784 & 0.2211 & 0.1201  \\

PRME
& 0.1159 & 0.2236 & 0.3105 & 0.3643 & 0.1712 
& 0.1052 & 0.2278 & 0.2944 & 0.3560 & 0.1786 
& 0.0521 & 0.1034 & 0.1425 & 0.1954 & 0.1002  \\
 
ST-RNN 
& 0.1483 & 0.2923 & 0.3622 & 0.4502 & 0.2198 
& 0.1409 & 0.3022 & 0.3577 & 0.4753 & 0.2212 
& 0.0799 & 0.1423 & 0.1940 & 0.2477 & 0.1429 \\

STGN 
& 0.1716 & 0.3381 & 0.4122 & 0.5017 & 0.2598 
& 0.1689 & 0.3391 & 0.3848 & 0.4514 & 0.2422 
& 0.0810 & 0.1842 & 0.2579 & 0.3095 & 0.1675 \\

STGCN 
& 0.1799 & 0.3425 & 0.4279 & 0.5214 & 0.2788 
& 0.1716 & 0.3453 & 0.3927 & 0.4763 & 0.2504 
& 0.0961 & 0.2097 & 0.2613 & 0.3245 & 0.1712 \\

PLSPL 
& 0.1917 & 0.3678 & 0.4523 & 0.5370 & 0.2806 
& 0.1889 & 0.3523 & 0.4150 & 0.4880 & 0.2542 
& 0.1072 & 0.2278 & 0.2995 & 0.3401 & 0.1847 \\

STAN 
& 0.2231 & 0.4582 & 0.5734 & 0.6328 & 0.3253 
& 0.1963 & 0.3798 & 0.4464 & 0.5119 & 0.2852 
& 0.1104 & 0.2348 & 0.3018 & 0.3502 & 0.1869 \\

\cmidrule(lr){1-6}
\cmidrule(lr){7-11}
\cmidrule(lr){12-16}

Ours 
& \textbf{0.2435} & \textbf{0.5089} & \textbf{0.6143} & \textbf{0.6880} & \textbf{0.3621} 
& \textbf{0.2254} & \textbf{0.4417} & \textbf{0.5287} & \textbf{0.5829} & \textbf{0.3262 }
& \textbf{0.1357} & \textbf{0.2852} & \textbf{0.3590} & \textbf{0.4241} & \textbf{0.2103}  \\
\bottomrule
\end{tabular}
\end{table*}

\subsection{Results}
Table \ref{tbl:result} shows the performance comparison between our model and the baselines on three datasets. We report the top-1, top-5, top-10, top-20 accuracy and MRR. Generally speaking, all models perform better on NYC and TKY datasets than CA. The main reason is POIs in NYC and TKY are constrained inside a relatively small area (New York city and Tokyo city). But POIs in CA datasets spread across California and Nevada, result in a sparser dataset.

For all datasets, our model outperforms the baselines by large margins. For example, in NYC dataset, we achieve 24.35\% top-1 accuracy while the best baseline STAN is 22.31\%. On top-5 accuracy, our model gains about 11\% improvement compared with baselines, and on top-20 accuracy, a 8.7\% performance increase is recorded. Similar results are shown in TKY. 
%since those two datasets are both collected from FourSquare platform and relatively more dense than Gowalla-CA dataset. 
Moreover, state-of-the-art attend-based or LSTM-based recommendation models such as STAN, PLSPL, STGCN perform significantly better than traditional Markov-chain based or Matrix factorization based models like FPMC, PRME. For example, the top-1 accuracy of STAN is twice as much as FPMC, namely 22.31\% to 10.03\% on NYC.

CA dataset contains 9.9k POIs and 250k check-in records widely spread in an area over 400,000 $km^2$
where the TKY dataset contains 7.8k POIs with 361k check-ins squeezed in about 2,000 $km^2$. Because of the more seriously data scarcity problem in both the number of checkins and spatial sparsity of POIs, models usually cannot reach the same level of performance in CA as they are in NYC and TKY. More specifically, the top-1 accuracy of STAN on NYC is 22.31\% which drops to 11.04\% on CA. Similar patterns appear on our model as well. The top-1 accuracy of our model on CA is 13.57\%, which beats all baselines but much lower than it on NYC dataset.

\subsection{Inspecting the \Graph} % Cold-start Performance
Previous section demonstrates the overall performance of the proposed framework. 
%We believe the huge performance gain owing to the proposed \graph, comprehensive POI embedding, fused time-category embeddings and transformer for modeling sequential data. 
In this section, we elaborate on the advantages of the trajectory flow map.
%which is of the most important contribution in this paper. 
We show that our method mitigates the cold-start problem for inactive users and short trajectories compared with other methods. Then we remove the \graph and compare the performance by learning POI embeddings via embedding layers. At last, we describe the quantitative features of the graph built for NYC dataset.

\subsubsection{Inactive users and active users} 
%First of all, we define the inactive users, most active users and middle users. 
Given a dataset, after splitting it into train, validation, and test, we count the number of trajectories of each user in the train set only. We then mark the top 15\% as the most active users, the bottom 15\% as inactive users, and the rest are normal users. Next, we evaluate the model on the test set and record the performance of different user groups.

Take NYC dataset as an example. Out of 1,047 users in the training set, we identify 137 inactive users with less than 13 trajectories; 144 most active users with more than 41 trajectories, and 766 normal users. In general, inactive users do not have rich historical information making it hard to learn from the data.
%a personalized user embedding. 
%Consequently, models usually cannot perform very well on the top-1 accuracy because an underlying principle is machine learning models learn from the experience (data). If the data is scarce, of course the learning can be very difficult. 
To mitigate this challenge, we built the global \graph and try to improve the recommendation results of inactive users by leveraging other users' sequences.

The performance in different user groups of our model and the baseline is shown in Table \ref{tbl:cold_start_user}. Compared with the baseline model, our model achieves higher accuracy in all three user groups. For inactive users, the top-1 and top-10 accuracy of our model is 12.24\% and 43.94\% while corresponding accuracy of STGN is 9.06\% and 31.24\%. A huge improvement is also observed on top-5 accuracy and top-20 accuracy. %The performance gain for inactive users is more difficult than users with sufficient historical data. Because we can build a larger model with more capacity to capture the useful patterns from vast amounts of data and have better performance. However, using a larger model when data is not enough may exacerbate the over-fitting problem.

\begin{table}[!htbp]

\caption{Cold Start (due to inactive users) performance on NYC}
\label{tbl:cold_start_user}
%\vspace{-0.2cm}
\begin{center}
\begin{tabular}{llllll}
\toprule  
User Groups & Model & Acc@1 & Acc@5 & Acc@10 & Acc@20 \\
\hline 
Inactive    & STGN & 0.0926 & 0.2435 & 0.3124 & 0.3763 \\
Normal      & STGN & 0.1778 & 0.3576 & 0.4189 & 0.5125 \\
Very active & STGN & 0.1813 & 0.3832 & 0.4274 & 0.5398 \\
\hline
Inactive    & Ours & 0.1224 & 0.3471 & 0.4394 & 0.4529 \\
Normal      & Ours & 0.2421 & 0.4739 & 0.5422 & 0.6430 \\
Very active & Ours & 0.2692 & 0.5639 & 0.6995 & 0.7762 \\
\bottomrule
\end{tabular}
\end{center}
%\vspace{-0.2cm}
\end{table}

\subsubsection{Short trajectories and long trajectories}
Short trajectories present another challenge in next POI recommendation. The lack of information in a short-trajectory is reflected in its limited spatio-temporal contexts, especially when the number of known check-ins in a trajectory is merely one or two. We define the short trajectory dynamically according to the dataset. In particular, we order trajectories in the test sets by length  and label the top 15\% as long trajectories and bottom 15\% as short ones. %Noted, sometimes the trajectories at bottom 15\% or 16\% have exactly the same length, and we group them into the same category. 

%In NYC test set, there are 1566 trajectories. Among them, 602 examples are short trajectories. 250 examples longer than 15 are long trajectories 
Table \ref{tbl:cold_start_traj} shows the performance of our mdoel and STGN on three trajectory groups of NYC dataset. Our model does not show significant difference on recommendation performance between the short trajectory and long trajectory. For example, the top-1 accuracy is 21.86\% and 24.52\% on short and long trajectories, and the top-20 accuracy of short trajectories is 57.82\%. However, STGN, an extended LSTM model, is quite sensitive to trajectory length and perform unsatisfying on short trajectories with 7.23\% top-1 accuracy and 33.28\% top-20 accuracy.

\begin{table}[!htbp]
\caption{Cold Start (due to short trajectory) performance on NYC}
\label{tbl:cold_start_traj}
\begin{center}
\begin{tabular}{llllll}
\toprule  
Trajectory & Model & Acc@1 & Acc@5 & Acc@10 & Acc@20 \\
\hline 
Short trajs  & STGN & 0.0723 & 0.2117 & 0.2784 & 0.3328 \\
Middle trajs & STGN & 0.1921 & 0.3471 & 0.4397 & 0.5231 \\
Long trajs   & STGN & 0.1934 & 0.3516 & 0.4380 & 0.5322 \\
\hline
Short trajs  & Ours & 0.2186 & 0.4561 & 0.5269 & 0.5782 \\
Middle trajs & Ours & 0.2441 & 0.4927 & 0.5881 & 0.6519 \\
Long trajs   & Ours & 0.2452 & 0.5378 & 0.6698 & 0.7681 \\
\bottomrule
\end{tabular}
\end{center}
%\vspace{-0.2cm}
\end{table}

\subsubsection{Removing trajectory flow map}
%Previously, we compare the performance of our model with baselines in cold start scenarios. 
For comparison, we remove the \graph from our model and learn the POI representations by embedding layers to show the contributions of the global \graph. Table~\ref{tbl:cold_start_wo_graph} shows the performance of model without \graph on three user groups and three trajectory groups. Summarizing  Tables~\ref{tbl:cold_start_user}, \ref{tbl:cold_start_traj} and \ref{tbl:cold_start_wo_graph}, we observe that after removing \graph, the top-10 accuracy of inactive users drops from 42.11\% to 34.70\%, and performance drops from 52.06\% to 47.21\% for short trajectories. These experiments verify that global \graph benefits the top-$k$ accuracy of inactive users and short trajectories when $k$ is large. However, for cold start problem, the top-1 accuracy do not gain much improvement from \graph. This phenomenon aligns to our assumption that incorporating generic movement patterns most help with top-$k$ accuracy for sufficiently large $k$. 
%next POI based on current location may not be the best solution but certainly not bad. 
Thus, even the top-1 accuracy does not show much difference, the top-10 performance is indeed better.

\begin{table}[!htbp]
\caption{Performance of proposed model without \graph on NYC}
\label{tbl:cold_start_wo_graph}
\begin{center}
\begin{tabular}{lclll}
\toprule  
& Model & Acc@1 & Acc@5 & Acc@10  \\
\hline 
Inactive users    & Ours w/o graph & 0.1105 & 0.2983 & 0.3470  \\
Normal users      & Ours w/o graph & 0.2094 & 0.4228 & 0.5010  \\
Very active users & Ours w/o graph & 0.2488 & 0.5271 & 0.6815  \\
\hline
Short trajs       & Ours w/o graph & 0.1914 & 0.4059 & 0.4721  \\
Middle trajs      & Ours w/o graph & 0.2129 & 0.4490 & 0.5232  \\
Long trajs        & Ours w/o graph & 0.2012 & 0.4882 & 0.6186  \\
\bottomrule
\end{tabular}
\end{center}
%\vspace{-0.2cm}
\end{table}

\subsubsection{Analysis of  the trajectory flow map}
\begin{figure}
	\centering
	\includegraphics[width=0.8\linewidth]{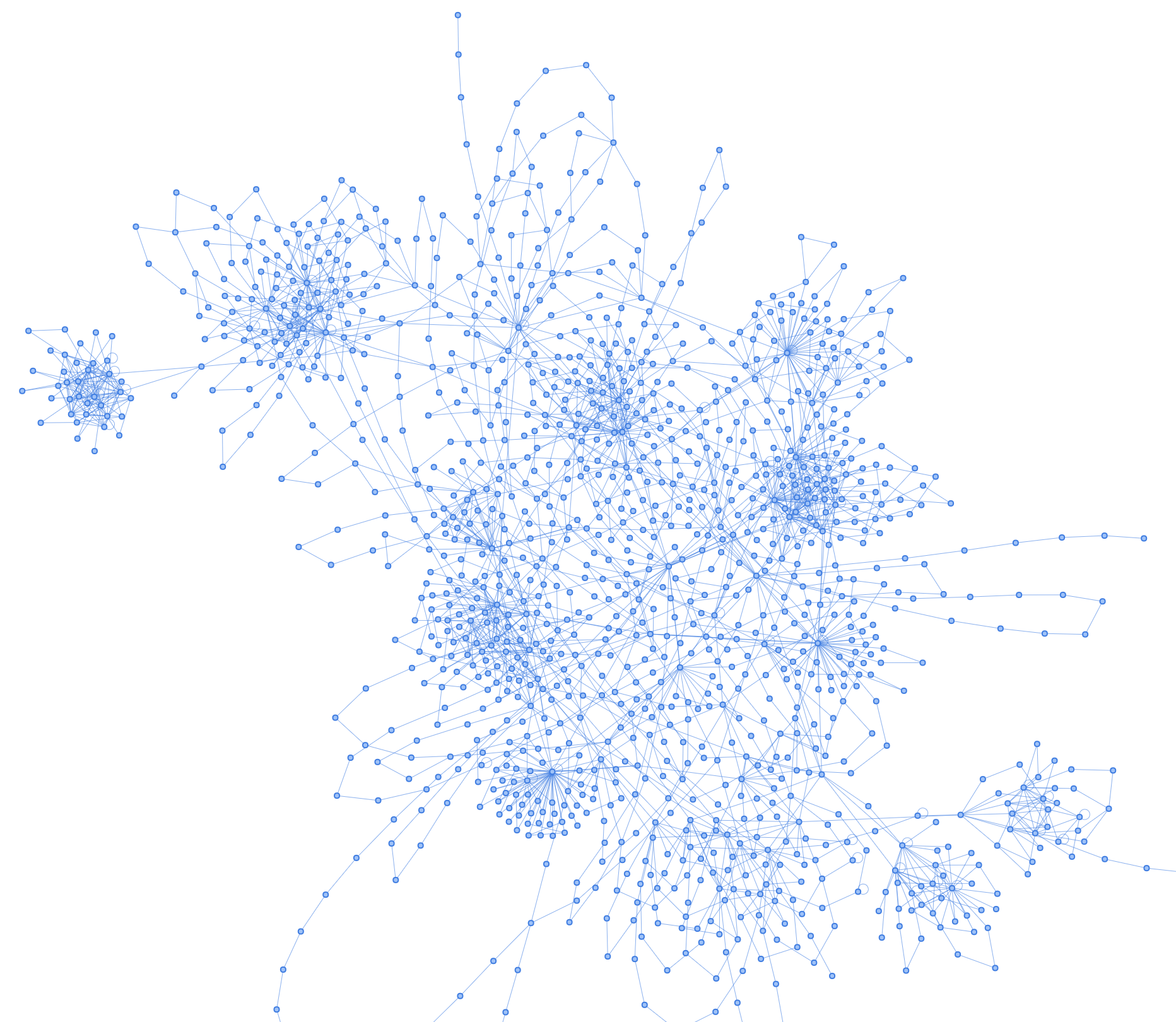}
	\caption{Partial \graph of NYC (directions and edge weights are removed for better visualization)}
	\label{fig:graph}
\end{figure}

%The trajectory flow map $\mathcal{G}$ is a weighted directed graph where nodes are POIs and edges represent the two successive check-ins happened between two POIs. The edge weight reflects the visiting frequency. 
We analyze structural properties of the trajectory flow map $\mathcal{G}$ constructed for NYC dataset as a case study. The constructed graph contains 4980 nodes and 37,756 edges. The average in-degree and out-degree are both 7.58. In a macroscopic perspective, there are about 7.58 possible choices for the next move in general. The mean edge weight is 1.91. That is to say if POI A and POI B are in two successive check-ins, then the average number of this check-in order is 1.91. Also, the average clustering coefficient of this graph is 0.1866, suggesting significant clustering. Figure~\ref{fig:graph} shows the partial graph of NYC dataset (without edge weight nor direction). The illustration demonstrates that the graph contains non-trivial structural information to be explored. In particular, the graph contains several visible communities, yet there are also many long paths suggesting that the graph deviates from a typical small-world network.

% print(nx.average_shortest_path_length(G)): Graph is not weakly connected.

% 讨论graph的avg path length, clustering coefficient等定量信息 (使用networkx)

\subsection{Ablation Study}
We conduct an ablation study to evaluate the impact of each proposed component to the model's final performance on NYC dataset. Particularly, seven experiments are conducted: 1) the full model; 2) without \graph and learn POI embeddings by embedding layers; 3) without transformer encoder and sequential data are handled by LSTM; 4) without time and category information; 5) without GCN and the random walk were employed to learn node embeddings from transtion graph; 6) instead of fusing user and POI embeddings, we concatenate them directly; 7) without time decoder and category decoder. In each experiment, rest settings remain unchanged as in the full model and only one component is either replaced or removed. The results are shown in Table \ref{tbl:ablation_study}.

\begin{table}[!htbp]
\caption{Ablation study: Comparing the full model with 6 variants}
\label{tbl:ablation_study}
\begin{tabularx}{\linewidth}{lXXXXX}
\toprule
 & Acc@1 & Acc@5 & Acc@10 & Acc@20 & MRR\\
\midrule
\textbf{Full Model} & \textbf{0.2435} & \textbf{0.5089} & \textbf{0.6143} & \textbf{0.6880} & \textbf{0.3621}  \\
w/o Graph & 0.2163 & 0.4512 & 0.5248 & 0.6402 & 0.3398  \\
w/o Transformer & 0.2221 &  0.4617 & 0.5520 & 0.6261 & 0.3317  \\
w/o Time\&Cat & 0.2296 & 0.4804 & 0.5489 & 0.6484 & 0.3495  \\
w/o GCN & 0.2062 & 0.4582 & 0.5601 & 0.6589 & 0.3187 \\
w/o Fusion & 0.2355 & 0.4731 & 0.5726 & 0.6614 & 0.3533 \\
Single decoder & 0.2207 & 0.4721 & 0.5722 & 0.6591 & 0.3443 \\ 
\bottomrule
\end{tabularx}
%\vspace{-0.1cm}
\end{table}

The full model achieves the best performance. For the rest components, the results suggest that \graph and GCN play a bigger role than other components in the overall performance. For example, top-1 accuracy drops from 24.35\% to 21.63\% and 20.62\% without \graph or using random walk to learn POI embeddings. Other components also contribute to the final recommendation results. For instance, the transformer can leverage the useful information from historical data automatically by the self-attention module, with less parameters to train and better performance.

%加上run time analysis

\section{Conclusion}
In this paper, we propose \GETNext, the first next POI recommendation model that utilizes graph-based learning on a global graph structure. In particular, we introduce \graph to capture generic movement patterns of the users to address the inactive user and short trajectory issues. We  defined sophisticated embeddings to encode spatio-temporal contexts, which include user-POI as well as time-category embeddings. We feed all embeddings into a transformer model, whose output is further enhanced through a transition attention map. Through a series of experiments on three real-world datasets, we demonstrate that our model significantly outperform all current state-of-the-art models by a large margin, and verify the benefits of the different components of our model. 
%
%and a Graph Enhanced Transformer (GETNext) for next POI recommendation. The \graph captures and represents the transitions between POIs as a graph. GETNext model leverage all trajectory flows from the graph along with other spatio-temporal contextual information to recommend the next possible POI and demonstrates outstanding performance. With the help of trajectory flow map, we can also alleviate two cold-start problems for next POI recommendation. 
For future work, we will further explore and distinguish the temporal patterns, e.g., between work-days and weekends. Another future work is to  classify users by their behaviors and build trajectory flow map for each type of users. 

\noindent \textbf{Acknowledgement.} This research is supported by the Marsden Fund Council from Government funding (MFP-UOA2123), administered by the Royal Society of New Zealand.

\bibliographystyle{ACM-Reference-Format}
\bibliography{reference}

% \appendix

\end{document}